\documentclass{ceab}  

\usepackage{epsfig}    
\usepackage{graphicx}   

\usepackage{ceabbib}     

\usepackage[T1]{fontenc}

\begin{document}

\title{VARIATIONS OF SOLAR NON-AXISYMMETRIC ACTIVITY}

\author{Gyenge, N., T. Baranyi and A. Ludm\'any
\vspace{2mm}\\
\it Heliophysical Observatory, Research Centre for Astronomy and Earth Sciences, \\
 Hungarian Academy of Sciences, Debrecen, P.O.Box 30, H-4010, \\
\it H-4010 Debrecen, P.O.Box 30. Hungary}

\maketitle

\begin{abstract}

The temporal behaviour of solar active longitudes has been examined by using two sunspot catalogues, the Greenwich Photoheliographic Results (GPR) and the Debrecen Photoheliographic Data (DPD). The time-longitude diagrams of the activity distribution reveal the preferred longitudinal zones and their migration with respect to the Carrington frame. The migration paths outline a set of patterns in which the activity zone has alternating prograde/retrograde angular velocities with respect to the Carrington rotation rate. The time profiles of these variations can be described by a set of successive parabolae. Two similar migration paths have been selected from these datasets, one northern path during cycles 21\,--\,22 and one southern path during cycles 13\,--\,14, for closer examination and comparison of their dynamical behaviours. The rates of sunspot emergence exhibited in both migration paths similar periodicities, close to 1.3 years. This behaviour may imply that the active longitude is connected to the bottom of convection zone.

\end{abstract}


\def\gore{active longitudes}

\section{Introduction}

The probability of magnetic field emergence is not uniform at all solar longitudes. Numerous works have been devoted to identify and follow the most active longitudinal belts but the different datasets, time intervals, approaches and preassumptions resulted in a broad variety of spatial patterns and temporal behaviours. In the history of this research field two larger groups can be separated. One of them disregards the spatially resolved active region data and focuses on the temporal variation of the activity by assuming that the active zones belong to a frame having different rotation rate from that of the Carrington frame. This approach is restricted by the preassumption that this different rotation rate is constant (Bai, 1987;  Balthasar, 2007;  Bogart, 1982;  Jetsu et al., 1997;  Olemskoy and Kitchatinov, 2007). The other group uses the position data of active regions but also with restricting preassumptions allowing e.g. the impact of differential rotation which necessarily implies a cyclic behaviour (Berdyugina and Usoskin, 2003; Usoskin et al., 2005; Zhang et al., 2011).

Our work tries to combine these two approaches. At first we want to identify the active belts and its motion with respect to the Carrington frame without assuming that this motion is invariable or cycle dependent. In the next phase the study of the temporal variation can focus on the moving active longitudinal zone. 

Our previous paper (Gyenge et al., 2012, hereinafter Paper I) presented a possible method to localise and follow the longitudinal zone of enhanced activity. The diagrams of that paper show a characteristic migration pattern in the time-longitude diagram. In contrast to the well recognizable Sp\"orer pattern in the time-latitude diagram, the longitudinal migration of enhanced activity does not show connection with the cycle profiles. A parabola has been fitted to the migration path and along this curve the width of the active zone was about 30 degrees, the flip-flop phenomenon was well identifiable. The migration path comprised the decreasing phase of cycle 21, the entire cycle 22 and the beginning of cycle 23 so it is highly improbable that the phenomenon of active longitude is connected with the solar cycle or the differential rotation. That work was restricted to the time interval of the Debrecen Photoheliographic Data (DPD): 1979-2011. The aim of the recent work is to extend the study to earlier cycles and to scrutinize closer the dynamics of flux emergence within the activity belt.

\section{Time-Longitude Analysis of Sunspot Distributions}

The work is based on the two detailed sunspot catalogue, the DPD (Gy\H ori et al., 2011) and the GPR (Greenwich Photoheliographic Results, Royal Observatory Greenwich). The procedure was similar to that of Paper I. The $360^{\circ}$ longitudinal circumference of the Sun was divided into $10^{\circ}$ bins and the normalized weight of activity has been computed in each bin and each Carrington rotation by the formula:

\begin{eqnarray}
      W_{i} = \frac{A_{i}}{ \sum_{j=1}^{36} A_{j} }
\end{eqnarray}

\begin{figure}[h!]
  \begin{center}
   \epsfig{file=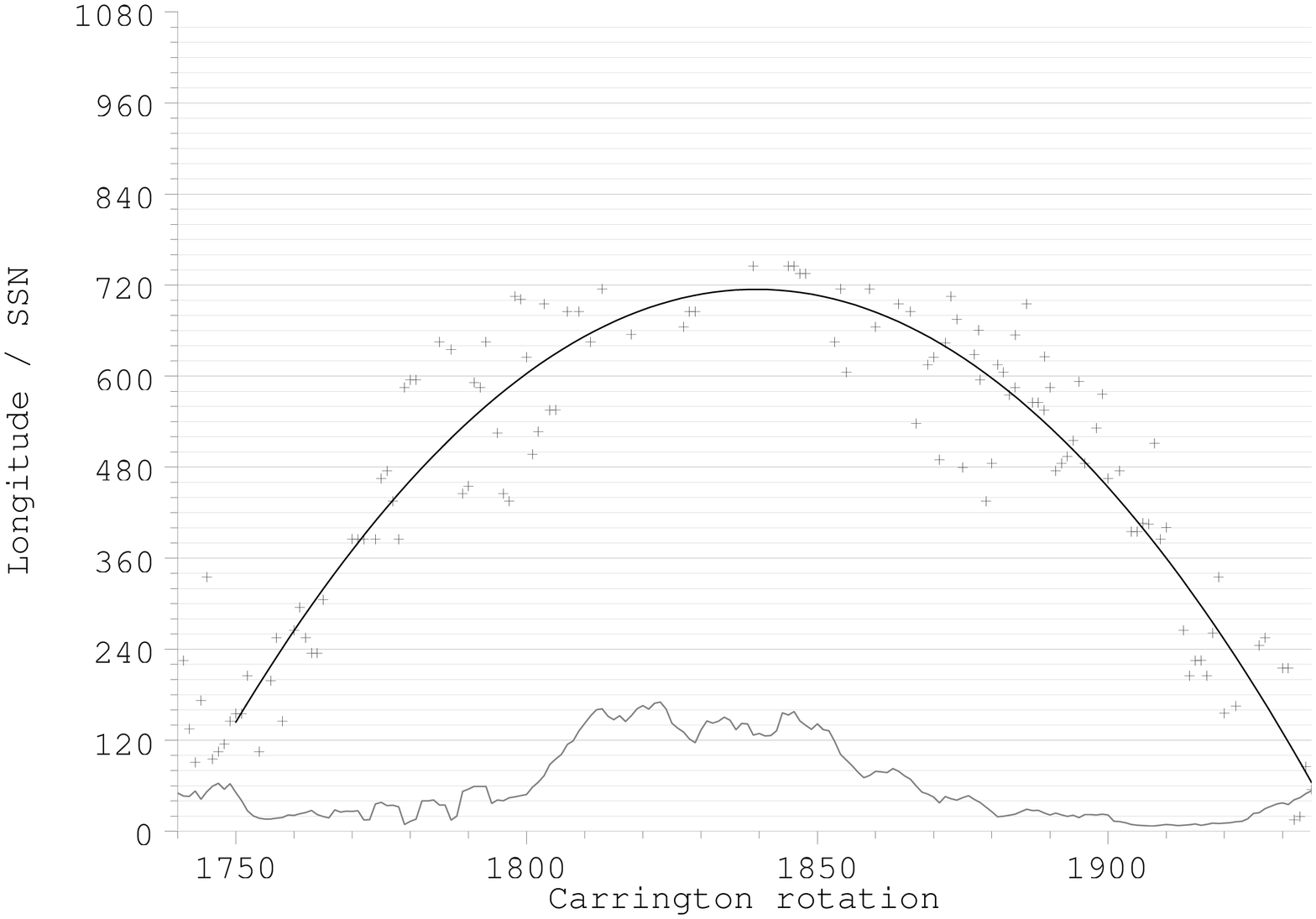,width=6.4cm,angle=-90}
   \epsfig{file=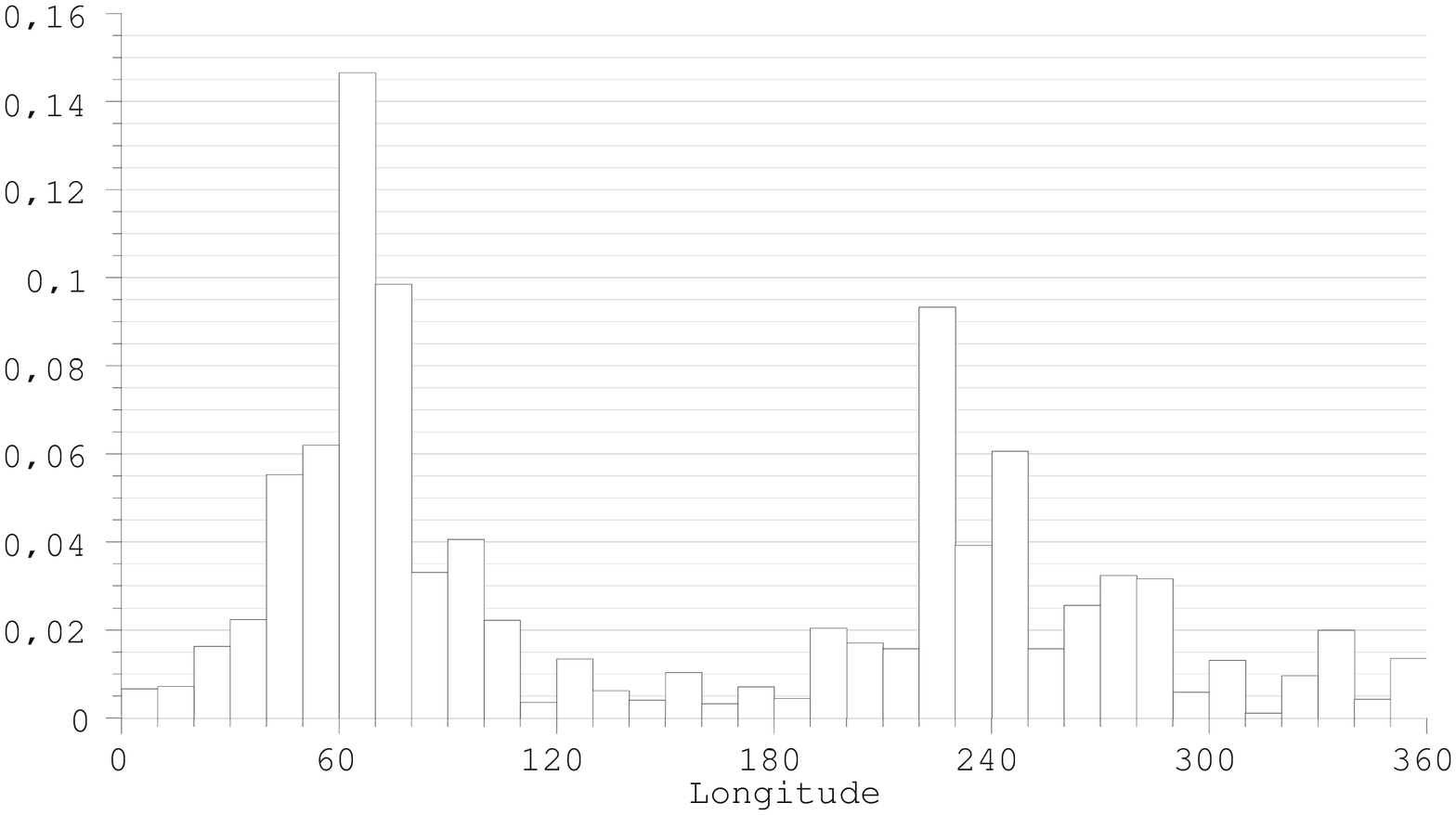,width=4.8cm,angle=-90}
  \end{center}
  \caption{Upper panel: migration of the active longitudinal zone with respect to the Carrington frame in the northern hemisphere between Carrington rotations 1740\,--\,1935 (between years 1984\,--\,1996). On the vertical axis the solar circumference is plotted three times. The lowest part of the panel contains the simultaneous cycle profiles plotted by using the International Sunspot Number (SIDC-team).  Lower panel: longitudinal distribution of activity as measured from the parabola.}
\end{figure} 

The $W_{i}$ quantity represents the fraction of the total activity emerging at a certain longitudinal bin so if its highest values are tracked through the rotations it may reveal the migration of the most active longitudinal zone with respect to the Carrington frame. The first step in the identification of the path was a search by visual pattern recognition in the time\,--\,longitude diagram, this was a forward and backward shift of the active zone between the Carrington rotations 1740\,--\,1935 (between years 1984\,--\,1996) in the northern hemisphere. This subjective choice has been checked with more objective procedures: fitting of a parabola on the points of the highest activity, the determination of the width of active zone along the parabola path, the detection of a flip-flop phenomenon. In the present work this procedure has been repeated with a somewhat different approach: in each Carrington rotation all longitudinal bins were disregarded in which the fraction of entire activity in the given rotation was smaller than 0.28 which is twice as high as the mean standard deviation of the averaged $W_{i}$ in the rotations. The remaining bins show up the migration path of the active zone, the parabola fitted on them is practically the same as in Paper I:

\begin{eqnarray}
	l=-0.081(r-1837)^{2}+720  
\end{eqnarray}

Where {\it l} is the longitude, and {\it r} is the Carrington rotation number. 

\begin{figure}[h!]
  \begin{center}
   \epsfig{file=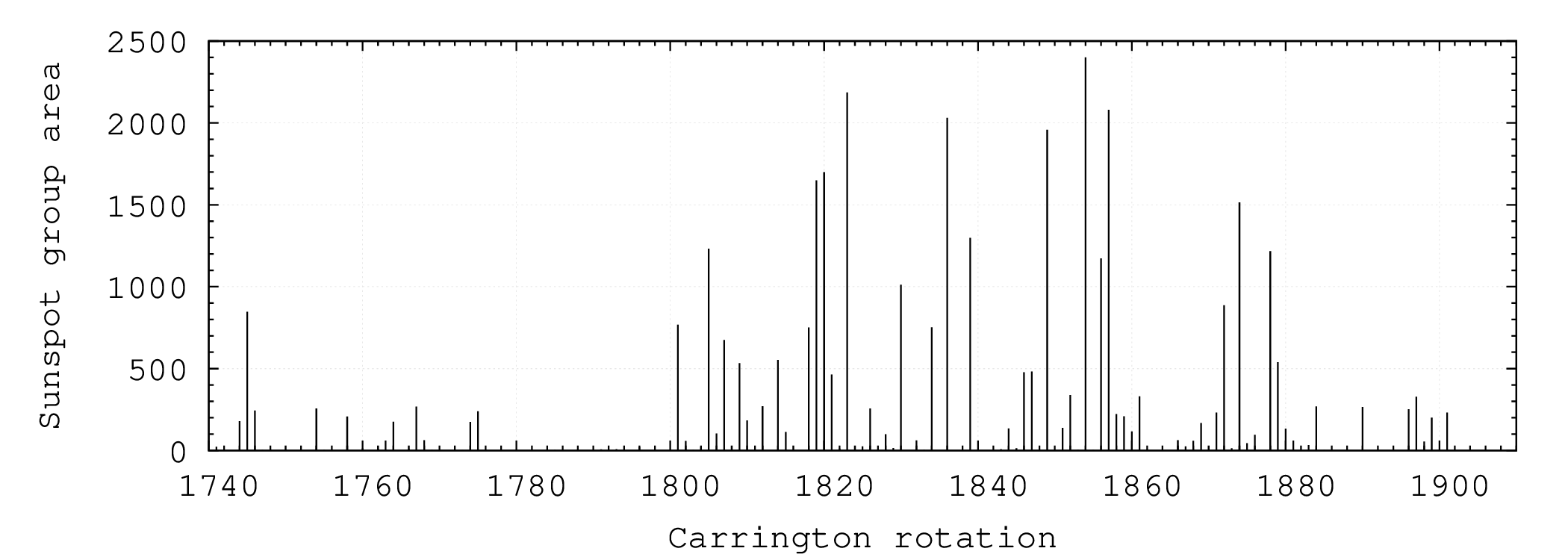,width=8cm}
   \epsfig{file=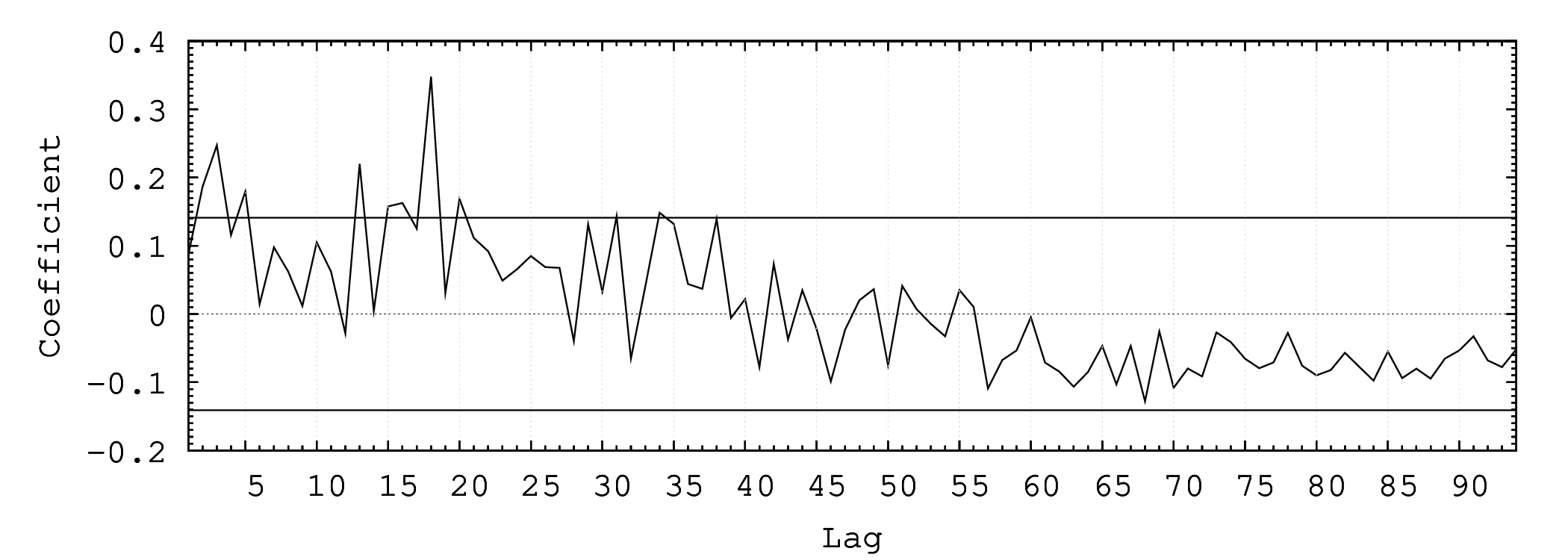,width=8cm}
   \epsfig{file=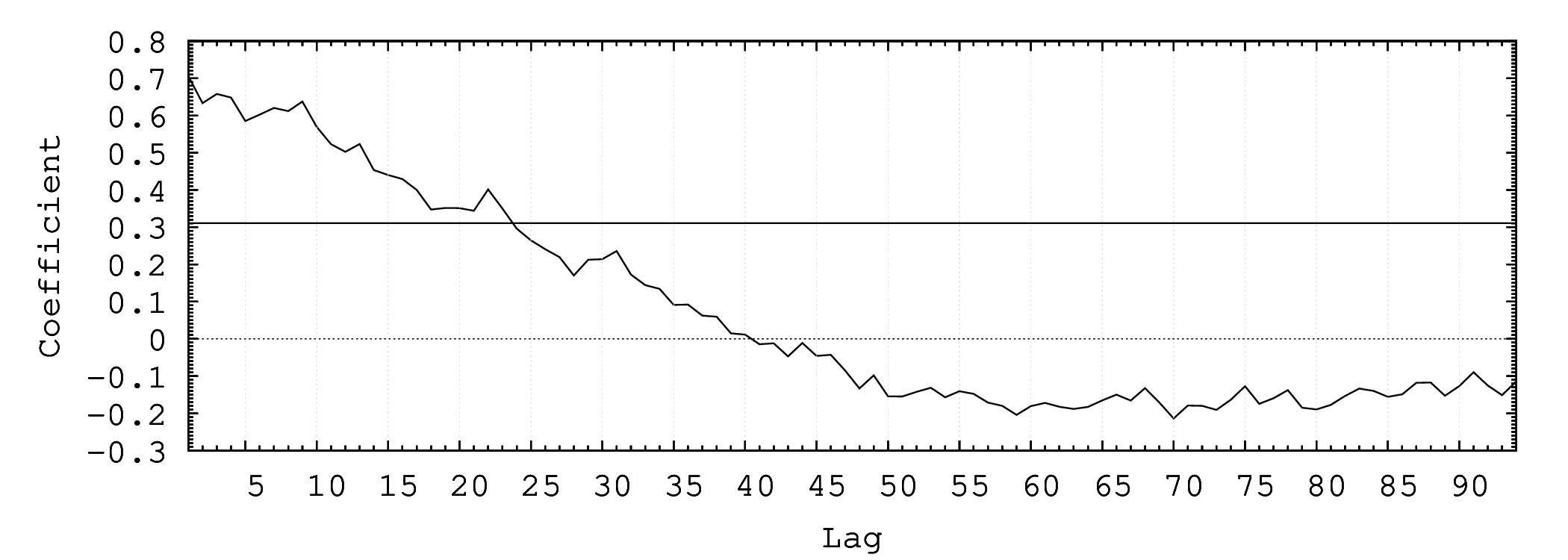,width=8cm}
  \end{center}
  \caption{Upper panel: temporal variation of monthly total sunspot area in the migrating active zone. Middle panel: the autocorrelogram of this data series. Lower panel: the autocorrelogram of the entire activity in the same time interval.}
\end{figure}

Figure 1 shows the path of the active zone, the selected points of the time\,--\,longitude diagram and the fitted parabola. The solar longitudinal circumference is plotted three times on the vertical axis in order to follow the migration. It is remarkable that the forward motion in the Carrington frame starts in the decreasing phase of cycle 21, it returns at the time of maximum of cycle 22 and it ends during the rising phase of cycle 23. The lower panel shows the longitudinal distribution of the activity in a moving reference frame in which the position of $60^{\circ}$ mark moves along the parabola of the migration path. The distributions in each Carrington rotation were averaged for the total length of the path. The activity distribution has a smaller secondary maximum at the opposite longitude of the main maximum.

\begin{figure}[h!]
  \begin{center}
   \epsfig{file=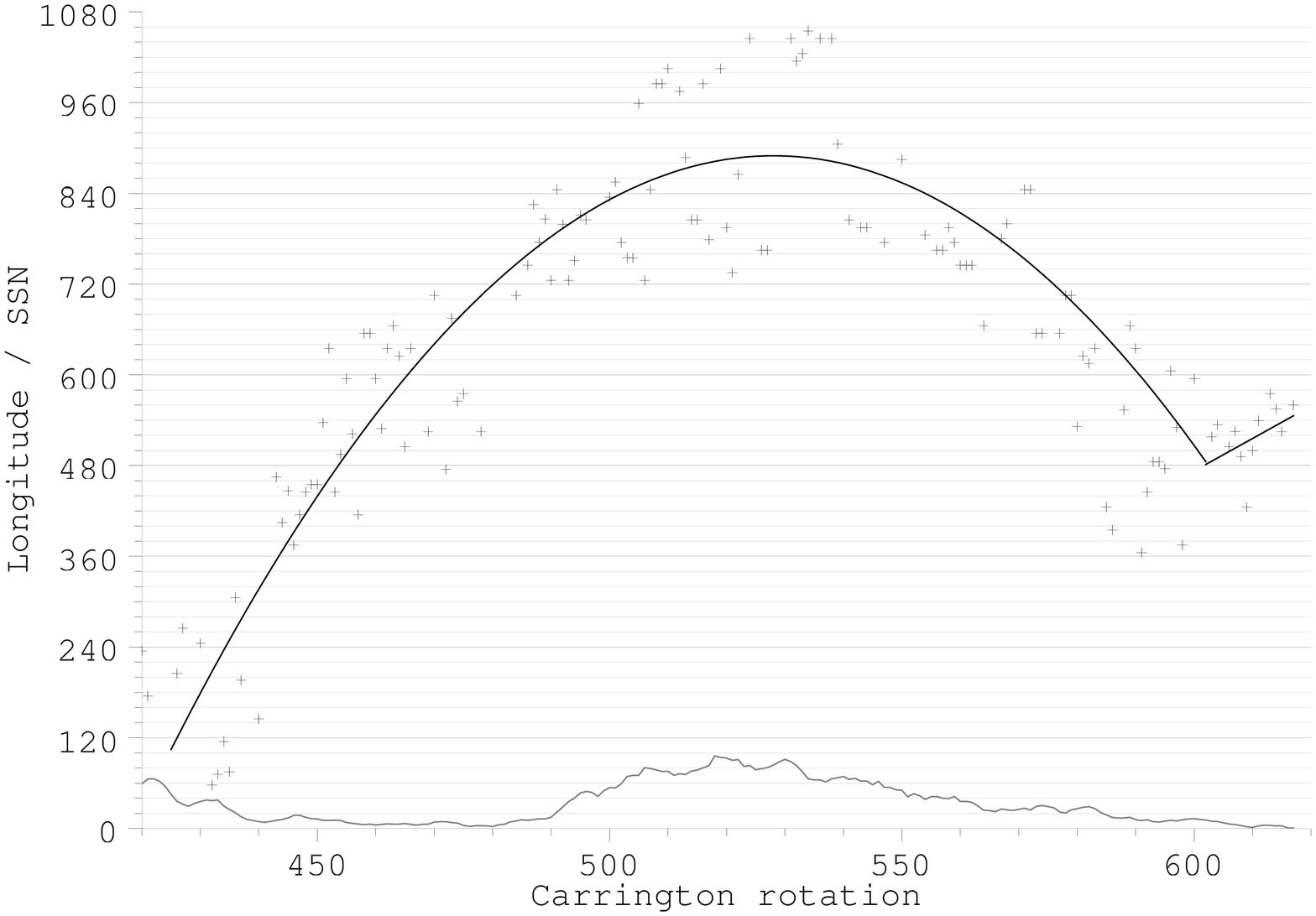,width=6.4cm,angle=-90}
   \epsfig{file=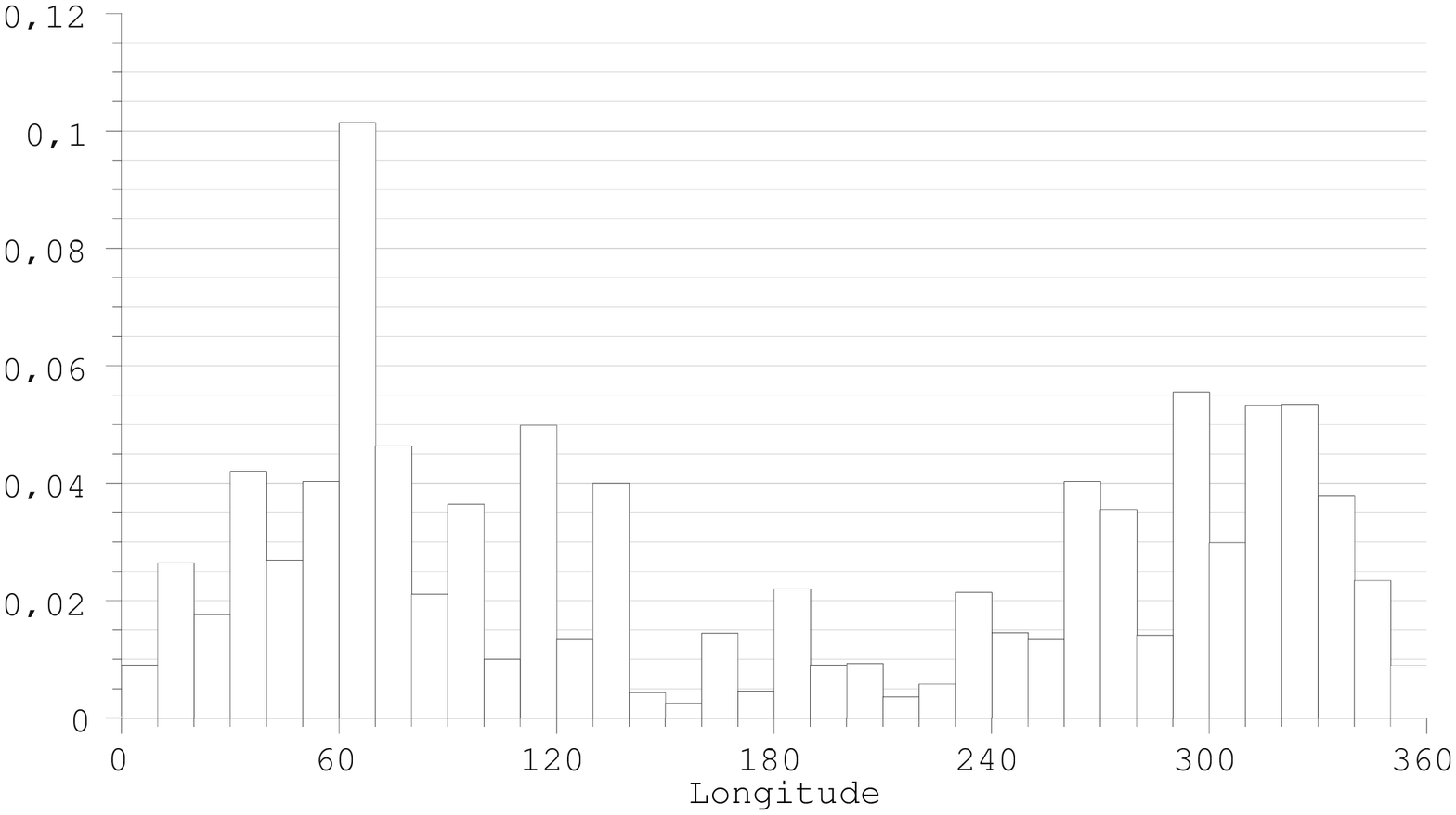,width=4.8cm,angle=-90}
  \end{center}
  \caption{The same diagrams as in Figure 1 for the migration path between Carrington rotations 420\,--\,620 (years 1885\,--\,1900)  }
\end{figure}

The temporal behaviour of the active zone was investigated by using autocorrelation analysis. Figure 2 shows the temporal variation and the autocorrelogram of monthly total sunspot area in the longitudinal zone of $\pm 15^{\circ}$ width on both sides of the parabola curve. The highest peak of the curve is at the rotation 18 which corresponds nearly to 1.3 years. To check whether this period belongs really to the active zone the autocorrelation of the entire activity has also been computed, see the lower panel of Figure 1, it does not contain this peak or some other signatures of any periodicity.

\begin{figure}[h!]
  \begin{center}
  \epsfig{file=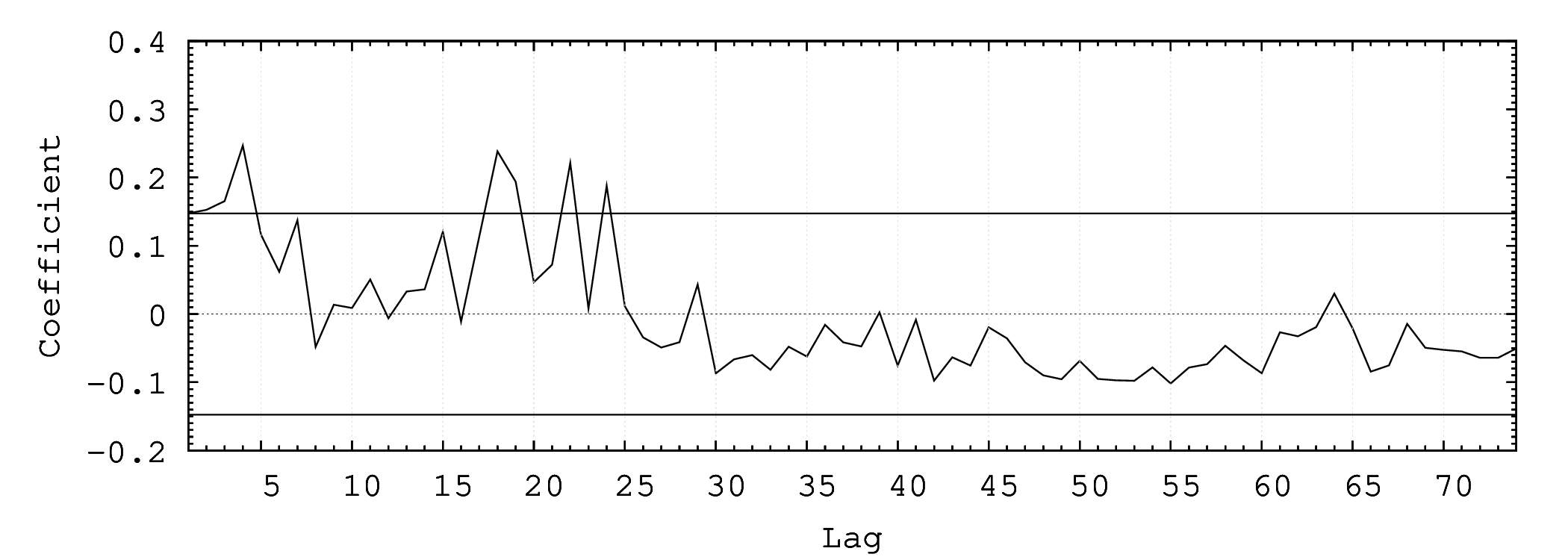,width=8cm}
  \epsfig{file=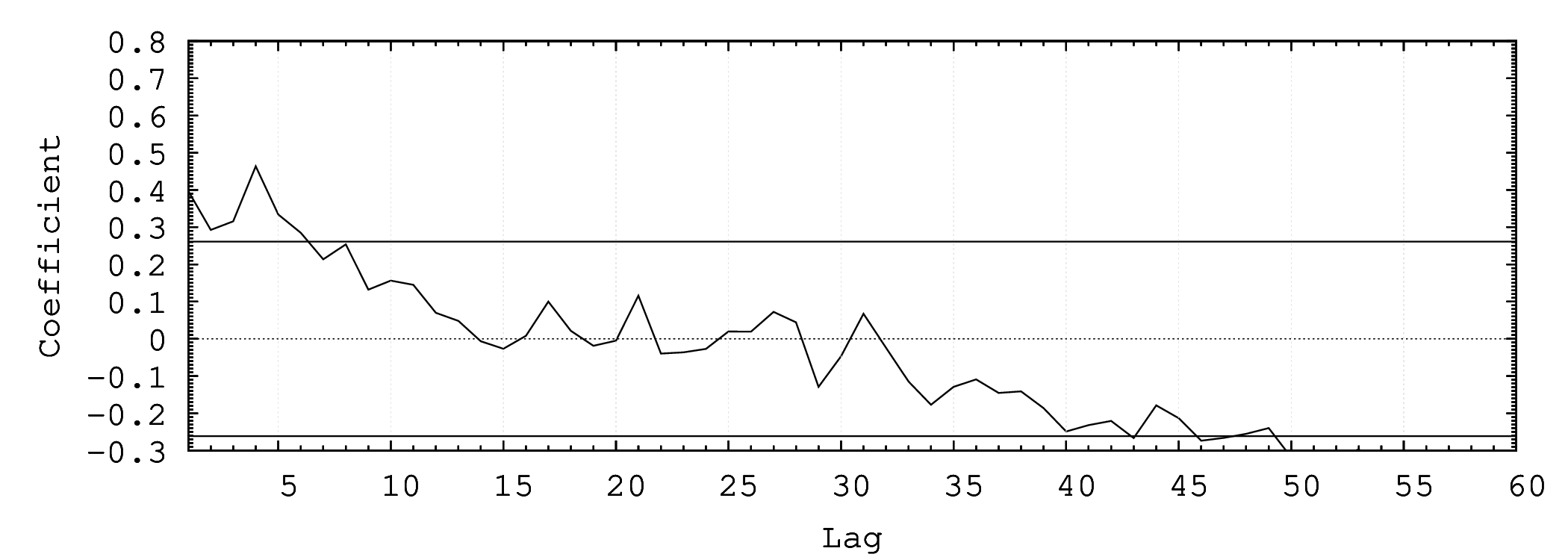,width=8cm}
  \end{center}
  \caption{The same diagrams as in the two lower panels of Figure 2 for the path shown in Figure 3. }
\end{figure}

The above presented investigations have been carried out by using the data of DPD. In order to extend the investigations the time-longitude diagrams of the GPR-period have been studied and another similar migration path has been selected in the southern hemisphere at the time of cycle 14 between Carrington rotations 420\,--\,620 (years 1885\,--\,1900). The procedure of curve fitting was the same as in the case of the path presented above. The obtained curve and the longitudinal activity distribution around it are plotted in Figure 3. 

There are remarkable similarities between these diagrams and those of Figure 1, primarily the parabola shape of the migration, the return of the migration at the time of cycle 14 maximum, and the main and secondary maxima at opposite positions in the longitudinal distribution of the activity obtained along the path. The most important difference is that the lengths of the forward-backward migrations are not the same, at around the rotation No. 600 a new migration path starts in forward direction, not followed here.

The most intriguing property of the northern migration path between 1984\,--\,1996 is that the activity within its narrow belt exhibits a variation with a period of about 1.3 years and this period cannot be detected in the entire activity of the same time interval. The same variation has also been studied in the migration path coinciding with cycles 13-14, the results are shown in Figure 4. In the autocorrelogram of Figure 4 the significant peak also exists at rotation 18 (1.33 years) but two smaller peaks are also present at rotations 22 and 24 (1.62 and 1.77 years). The check of the curve, the autocorrelogram of the entire activity is similar to that of the studied interval at cycles 21\,--\,23, it has no significant peaks at all.

\section{Discussion}

The obtained results of the above case studies may have connections to several other findings. The shapes of the migration paths of the active longitudinal zones are fairly similar to those variations which have been found by Juckett (2006) with a quite different method. The variations of the flux emergences within the active zones exhibit a period of 1.3 years, the active zone during cycle 14 has also two further periods of smaller peaks. These periods are absent in the entire activity. This may imply that the magnetic fluxes of these active zones emerge from the bottom of the convective zone where the 1.3 year radial torsional oscillation has been detected (Howe et al, 2000). This interpretation is supported by the theoretical considerations of Bigazzi and Ruzmaikin (2004) that the active longitude can only be pertinent at the bottom of the convection zone because at higher layers the differential rotation would disarrange it. If the conjectured connection to the depth of the tachocline zone really exists it could support the "shallow layer" model of the active longitudes proposed by Dikpati and Gilman (2005). 

The present work studied a northern and a southern migration path at an interval of about a century from each other and several similarities have been found between them. A large statistical study is in preparation for the detailed search for further identifiable migrating active longitudes in the entire time interval covered by the DPD and GPR.

\section*{Acknowledgements} 
The research leading to these results has received funding from the European Community's Seventh Framework Programme (FP7/2012-2015) under grant agreement No. 284461.


\section*{References}
\begin{itemize}
\small
\itemsep -2pt
\itemindent -20pt

\item[] Bai, T., 1987, {\it \apj}, 314, 795-807.
\item[] Balthasar, H., 2007, {\it \aap} 471, 281-287.
\item[] Berdyugina, S. V. \& Usoskin, I. G., 2003, {\it \aap} 405, 1121-1128.
\item[] Bigazzi, A. \& Ruzmaikin, A., 2004,  {\it \apj}, 604, 944-959.
\item[] Bogart, R. S., 1982, {\it \solphys}, 76, 155-165.  
\item[] Dikpati, M. \& Gilman, P., 2005, {\it \apj}, 635, L193-L196.
\item[] Gyenge, N., Baranyi, T. \& Ludm\'any, A., 2012, {\it CEAB}, 36, No.1., 9-16.  Paper I.
\item[] Gy\H ori, L., Baranyi, T., \& Ludm\'any, A. 2011, {\it IAUS} 273, 403. \\ see: http://fenyi.solarobs.unideb.hu/DPD/index.html
\item[] Howe, R., Christensen-Dalsgaard, J., Hill, F., et al., 2000, {\it Science}, 287, 2456-2460.  
\item[] Jetsu, L., Pohjolainen, S., Pelt, J., \& Tuominen, I., 1997, {\it \aap}, 318, 293-307.  
\item[] Juckett, D. A., 2006, {\it \solphys}, 245, 37-53.
\item[] Olemskoy, S. V. \& Kitchatinov, L. L., 2007, {\it Geomagn.Aeron.}, 49, 866-870. 
\item[] Royal Observatory Greenwich, Greenwich Photoheliographic Results, 1874-1976, in 103 volumes, see: http://solarscience.msfc.nasa.gov/greenwch.shtml
\item[] SIDC-team, World Data Center for the Sunspot Index, Royal Observatory of Belgium, Monthly Report on the International Sunspot Number, online catalogue of the sunspot index: http://www.sidc.be/sunspot-data/
\item[] Usoskin, I. G., Berdyugina, S. V. \& Poutanen, J., 2005, {\it \aap}, 441, 347-352.
\item[] Zhang, L.Y., Mursula, K., Usoskin, I. G. \& Wang, H.N., 2011, {\it \aap}, 529, 23.

\end{itemize}

\end{document}